\documentclass[sigconf]{acmart}

\AtBeginDocument{%
 }
  
\copyrightyear{2024}
\acmYear{2024}
\setcopyright{rightsretained}
\acmConference[CIKM '24]{Proceedings of the 33rd ACM International Conference on Information and Knowledge Management}{October 21--25, 2024}{Boise, ID, USA}
\acmBooktitle{Proceedings of the 33rd ACM International Conference on Information and Knowledge Management (CIKM '24), October 21--25, 2024, Boise, ID, USA}
\acmDOI{10.1145/3627673.3679717}
\acmISBN{979-8-4007-0436-9/24/10}

\usepackage{bbm}
\usepackage{balance}
\usepackage{cleveref}
\usepackage{multirow}
\usepackage{graphicx}
\usepackage{enumitem}
\usepackage{xcolor}

\usepackage{algorithm,algpseudocode}%

\usepackage{bm}
\usepackage{color}
\usepackage{apxproof}
\usepackage{gensymb}
\usepackage{pifont}%

\usepackage{xspace}
\usepackage{mathtools}

\newcommand{\eg}{\emph{e.g.}}
\newcommand{\ie}{\emph{i.e.}}

\begin{document}

\newcommand{\ours}{{RecBLR}\xspace}

\title{Behavior-Dependent Linear Recurrent Units for Efficient Sequential Recommendation}

\author{Chengkai Liu}
\affiliation{
  \institution{Texas A\&M University}
  \city{College Station, TX}
  \country{USA}}
\email{liuchengkai@tamu.edu}

\author{Jianghao Lin}
\affiliation{
  \institution{Shanghai Jiao Tong University}
  \city{Shanghai}
  \country{China}}
\email{chiangel@sjtu.edu.cn}

\author{Hanzhou Liu}
\affiliation{
  \institution{Texas A\&M University}
  \city{College Station, TX}
  \country{USA}}
\email{hanzhou1996@tamu.edu}

\author{Jianling Wang}
\affiliation{
  \institution{Texas A\&M University}
  \city{College Station, TX}
  \country{USA}}
\email{jwang713@tamu.edu}

\author{James Caverlee}
\affiliation{
  \institution{Texas A\&M University}
  \city{College Station, TX}
  \country{USA}}
\email{caverlee@cse.tamu.edu}

\renewcommand{\shortauthors}{Liu et al.}

\begin{abstract}
Sequential recommender systems aims to predict the users' next interaction through user behavior modeling with various operators like RNNs and attentions. 
However, existing models generally fail to achieve the three golden principles for sequential recommendation simultaneously, \ie, training efficiency, low-cost inference, and strong performance.
To this end, we propose \textbf{\ours}, an Efficient Sequential \underline{Rec}ommendation Model based on \underline{B}ehavior-Dependent \underline{L}inear \underline{R}ecurrent Units to accomplish the impossible triangle of the three principles. By incorporating gating mechanisms and behavior-dependent designs into linear recurrent units, our model significantly enhances user behavior modeling and recommendation performance. 
Furthermore, we unlock the parallelizable training as well as inference efficiency for our model by designing a hardware-aware scanning acceleration algorithm with a customized CUDA kernel. 
Extensive experiments on real-world datasets with varying lengths of user behavior sequences demonstrate \ours's remarkable effectiveness in simultaneously achieving all three golden principles - strong recommendation performance, training efficiency, and low-cost inference, while exhibiting excellent scalability to datasets with long user interaction histories.
\end{abstract}

\ccsdesc[500]{Information systems~Recommender systems}
\keywords{Recommender Systems, Sequential Recommendation}

\maketitle

\section{Introduction}

\begin{figure}[t]
    \centering
    \includegraphics[width=0.32\textwidth]{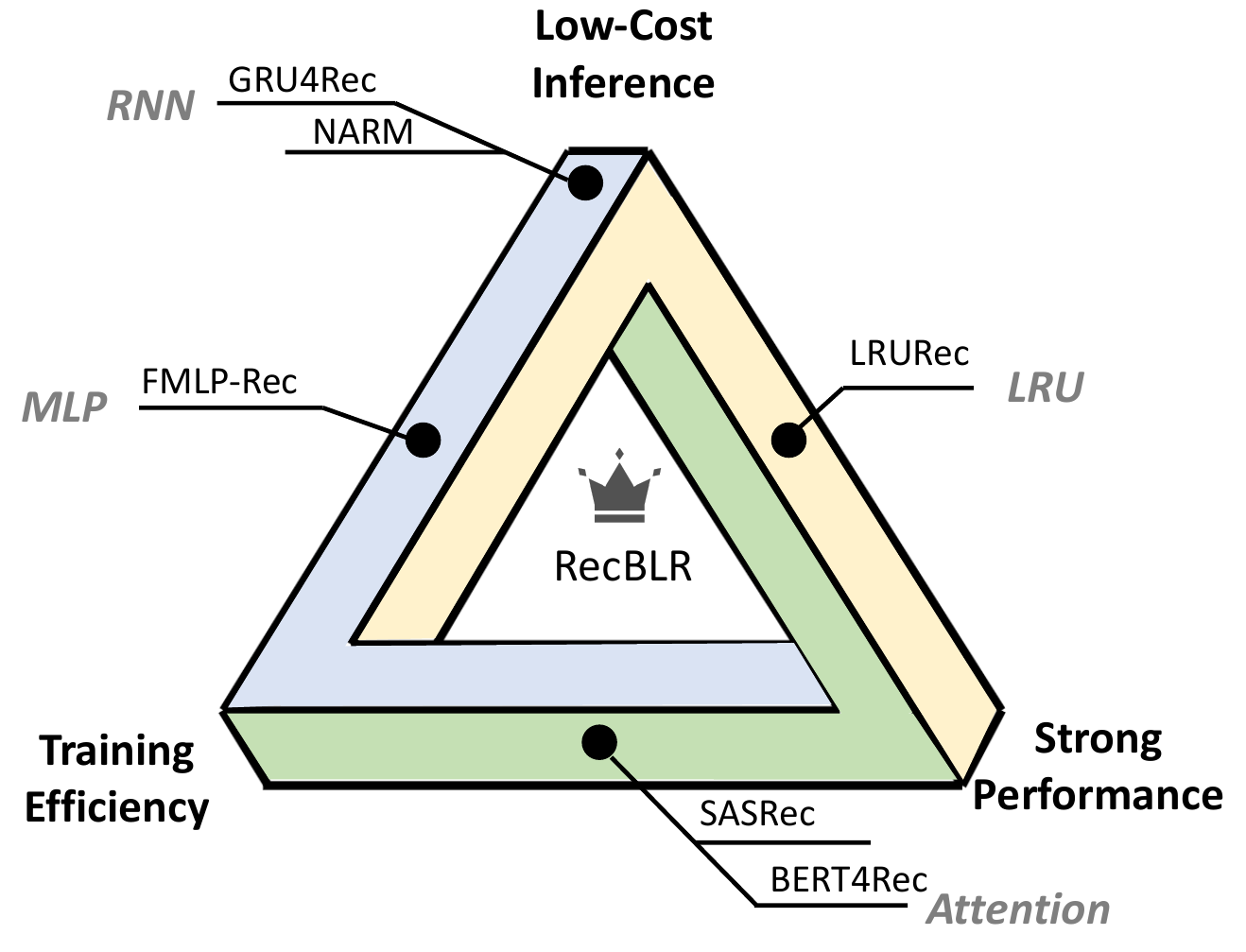}
    \caption{The triangle of three gold principles for the model design in sequential recommendation.}
    \vspace{-2pt}
    \label{fig:golden principle}
\end{figure}

Sequential recommender systems are crucial for personalized online services, capturing evolving user preferences through past interactions and predicting future actions by effectively modeling sequential dependencies within historical user behaviors~\cite{chen2018sequential,xie2022contrastive,liu2024mamba4rec,lin2024rella}. 
The core of sequential recommendation is the user behavior modeling techniques, which have evolved significantly over time, advancing from traditional methods such as Markov chains~\cite{he2016fusing,rendle2010factorizing} to sophisticated deep learning methods~\cite{hidasi2015session,yan2019cosrec}.
As illustrated in Figure~\ref{fig:golden principle}, there is a triangle of golden principles for designing the user behavior modeling operators for sequential recommendation, \ie, \textit{training efficiency}, \textit{low-cost inference}, and \textit{strong performance}.

As shown in Figure~\ref{fig:golden principle}, earlier approaches like GRU4Rec~\cite{hidasi2015session} and NARM~\cite{li2017neural} are built based on recurrent neural networks (RNNs)~\cite{chung2014empirical,sherstinsky2020fundamentals} or convolutional neural networks (CNNs)~\cite{o2015introduction,krizhevsky2012imagenet}. Although they possess the low-cost inference property, they can easily suffer from inferior recommendation performance, as well as training inefficiency caused by their non-parallelism natures~\cite{sun2023retentive}. 
As an alternative, FMLP-Rec~\cite{zhou2022filter} designs filtering-enhanced multi-layer perceptron (MLP) operators for user behavior modeling to achieve high efficiency in terms of both training and inference. 
However its performance is still limited due to the catastrophic forgetting issues~\cite{schak2019study, kirkpatrick2017overcoming}. 
Meanwhile, as the Transformer~\cite{vaswani2017attention} architecture is flourishing in natural language processing (NLP) domains, researchers start to investigate the potential of attention operators for sequential recommendation. 
Classical models like SASRec~\cite{kang2018self} and BERT4Rec~\cite{sun2019bert4rec} enjoy both training parallelism and impressive performance at the cost of inefficient inference, because of the quadratic computational complexity and memory-bound key-value
cache~\cite{shazeer2019fast}.
As the online platforms grow and user interaction sequences become longer, the limitation of attention mechanisms exacerbates, impeding scalability for sequential recommendation with long user behavior sequences. 
As the latest research, LRURec~\cite{yue2024linear} pioneers the application of linear recurrent unit (LRU) operators~\cite{martin2017parallelizing, orvieto2023resurrecting} for sequential recommendation. 
It achieves strong performance and low-cost inference, but turns out training inefficient due to the introduction of non-linearity and sequence padding.

Based on the discussions above, we can observe that the existing sequential recommendation models seemingly suffer from the ``impossible triangle'' of the three golden principles as illustrated in Figure~\ref{fig:golden principle}.
That is, none of they could achieve superior performance and meanwhile enjoy high efficiency in terms of both training and inference. 
Therefore, in this paper, we aim to propose and address such an open research question: \textbf{can we establish a model with delicately designed user behavior modeling operators that could achieve all the three golden principles for sequential recommendation?}

We choose the linear recurrent unit (LRU)~\cite{martin2017parallelizing, orvieto2023resurrecting} as the basic operator, and start by investigating its potential drawbacks in the case of sequential recommendation scenarios when considering the triangle of three golden principles. 
\textit{Firstly}, the LRU operator is independent to the input user behaviors, \ie, every single user behavior is transformed through the same network, thereby contributing equally to the user preference representation. 
However, the diverse behaviors in the user sequence should make different contributions due to the multifaceted nature of user interests. 
Hence, it is essential to build a behavior-dependent operator to improve the performance of user behavior modeling.
\textit{Secondly}, previous work, \ie, LRURec~\cite{yue2024linear}, simply follows the classical design of LRU by leveraging the eigenvalue decomposition in the complex domain. 
Nevertheless, as demonstrated in~\cite{gu2023mamba}, complex elements are beneficial for continuous modalities like video and audio, but not for discrete sequence modeling like user behaviors. 
Moreover, complex-domain LRUs introduce a larger number of parameters compared to real-valued counterparts, potentially leading to overfitting and degenerated performance. 
\textit{Lastly}, while the above two points both aim at the ``strong performance'' principle, they can largely hurt the training and inference efficiency due to the complicated design of behavior modeling operators (\ie, behavior-dependent and non-complex). 
Therefore, it is crucial to move beyond the simple model design, and consider hardware-aware parallelism acceleration algorithm for the rest two principles of ``training efficiency'' and ``low-cost inference'' simultaneously.

To this end, in this work, we propose \textbf{\ours}, an Efficient Sequential \underline{Rec}ommendation Model with \underline{B}ehavior-Dependent \underline{L}inear \underline{R}ecurrent Units. 
\ours first achieves the ``strong performance'' principle by discarding complex numbers and designing a novel Behavior-Dependent Linear Recurrent Unit (BD-LRU). 
BD-LRU, as the core of our user behavior modeling operators, introduces the gating mechanisms to capture the multifaceted characteristics of different user behaviors, thereby cast a multi-interest modeling over the whole user sequence.
Then, to compensate for the potential training overhead introduced by our BD-LRUs, we design a hardware-aware scanning acceleration algorithm for efficient parallelizable training. Specifically, we introduces a parallel scanning mechanism together with a customized CUDA kernel to significantly enhance the model's computational efficiency. 
Putting it all together, \ours is therefore specifically tailored for efficient sequential recommendation with superior performance, succeeding in achieving the triangle of three golden principles simultaneously, \ie, training efficiency, low-cost inference, and strong performance.

The key contributions of this paper are as follows:
\begin{itemize}
[leftmargin=*,noitemsep,topsep=1.5pt]
    \item We propose a novel Efficient Sequential \underline{Rec}ommendation Model with \underline{B}ehavior-Dependent \underline{L}inear \underline{R}ecurrent Units (\textbf{RecBLR}), which achieves the three golden principles simultaneously, \ie, training efficiency, low-cost inference, and strong performance.
    \item For \textit{strong performance}, we discard complex numbers and design the Behavior-Dependent Linear Recurrent Unit (BD-LRU) operator, which capture the difference among diverse behaviors and thereby leads to improved multi-interest user behavior modeling.
    \item For \textit{training efficiency} and \textit{low-cost inference}, we further design a hardware-aware scanning acceleration algorithm with a customized CUDA kernel to enable efficiency computation.
    \item We conduct extensive experiments on datasets with various sequence lengths, demonstrating the effectiveness, efficiency, and scalability of our model, particularly on datasets with long user interaction sequences.
\end{itemize}

\section{Preliminaries}

\begin{figure*}[ht]
    \vspace{-0.1in}
    \centering
    \includegraphics[width=0.75\textwidth]{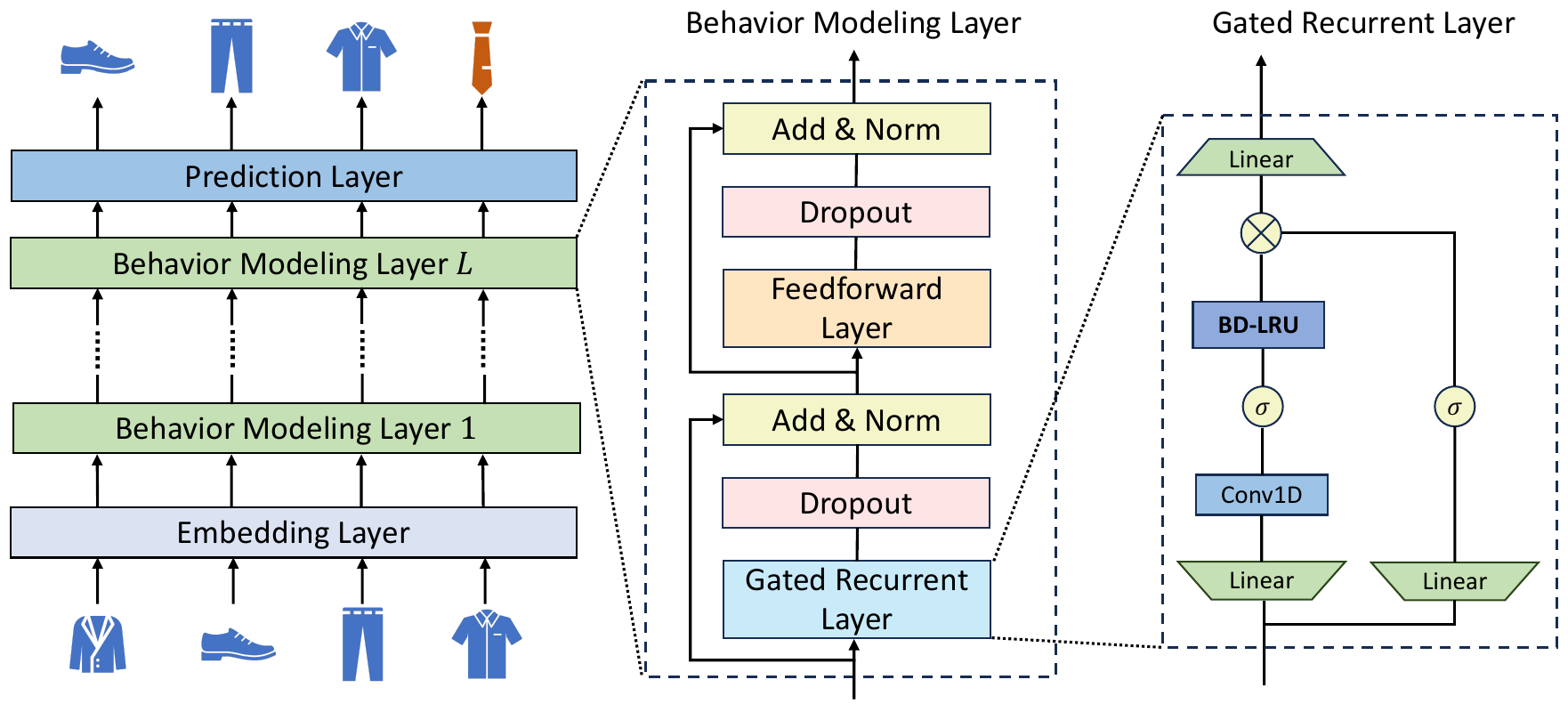}
    \vspace{-0.15in}
    \caption{The overall architecture of \ours.}
    \label{fig:overview}
    \vspace{-0.15in}
\end{figure*}

For sequential recommendation, we represent the user set as $\mathcal{U}=\left\{u_i\right\}_{i=1}^{|\mathcal{U}|}$, the item set as $\mathcal{V}=\left\{v_i\right\}_{i=1}^{|\mathcal{V}|}$, and the interaction sequence of a user $u \in \mathcal{U}$ as $\mathcal{S}_u=\left\{v_t\right\}_{t=1}^T$. This interaction sequence is chronologically ordered, and its length is denoted by $T$. Given a user's history $\mathcal{S}_u$, the task of sequential recommendation is to predict the next item $v_{T+1}$ that user $u$ is likely to interact with.

\subsection{Sequential Recommendation Models}

Mainstream neural sequential recommendation models are typically comprised of three key components: the embedding layer, behavior modeling layer, and prediction layer.

\noindent \textbf{Embedding Layer.}
Similar to existing sequential recommendation models, our model employs a standard embedding layer to map item IDs into a high-dimensional vector space. The embedding layer utilizes a learnable embedding matrix $\boldsymbol{E} \in \mathbb{R}^{|\mathcal{V}| \times D}$, where $\mathcal V$ denotes the item set and $D$ represents the embedding dimension. By applying the embedding layer to the input item sequence $\mathcal{S}_u$ of length $T$, along with dropout~\cite{srivastava2014dropout} and layer normalization~\cite{ba2016layer}, we obtain the item embedding $H \in \mathbb{R}^{T \times D}$:
\begin{equation}
 H =\operatorname {LayerNorm} (\operatorname {Dropout} ( \mathcal {S}_u \boldsymbol E)).
\label{eq:layernorm}
\end{equation}
While Transformer-based sequential recommendation models~\cite{kang2018self, sun2019bert4rec, yuan2022multi} often utilize positional embedding to capture chronological information, our recurrent architecture inherently captures position awareness. Therefore, we discard the need for additional positional embedding commonly used in the Transformer model~\cite{vaswani2017attention}.

\noindent \textbf{Behavior Modeling Layer.}
The primary distinction across various sequential recommendation models resides in the behavior modeling layer, using different operators such as attentions~\cite{kang2018self, sun2019bert4rec}, convolutional neural networks (CNNs)~\cite{tang2018personalized}, or recurrent neural networks (RNNs)~\cite{hidasi2015session, li2017neural}. In the following methodology section, we illustrate our proposed behavior modeling layer in detail.

\noindent \textbf{Prediction Layer.} 
Most neural sequential recommenders share a common prediction layer design. We leverage this standard prediction layer (including neural baselines in this paper), which utilizes the last item embedding to generate the final prediction scores:
\begin{equation}
\hat{y} = \operatorname{Softmax}(h_T\boldsymbol{E}^\top) \in \mathbb{R}^{|\mathcal{V}|},
\end{equation}
where $\boldsymbol{E}$ is the embedding matrix and $h \in \mathbb{R}^D$ is the last item embedding obtained from the final output of the behavior modeling layers. $\hat{y} \in \mathbb{R}^{|\mathcal{V}|}$ represents the probability distribution over the next item in the item set $\mathcal{V}$.

\subsection{Linear Recurrent Units}

Traditional Recurrent Neural Networks (\eg, LSTM~\cite{hochreiter1997long}, GRU~\cite{chung2014empirical}) have achieved remarkable success in many sequential tasks. However, their inherent non-linear dependencies between hidden representations hinder efficient parallelizable training. Linear Recurrent Neural Networks have emerged as a promising alternative, addressing this limitation by employing linear transformations instead of non-linear activation functions such as tanh or ReLU functions. To achieve parallelizable training, Linear Recurrent Units (LRUs) are introduced in~\cite{gu2021efficiently}: 
\begin{equation}
    \label{eq:lru}
    \begin{aligned}
    h_t & = A h_{t-1} + B x_t, \\
    y_t & = C h_t + D x_t,
    \end{aligned}
\end{equation}
where $A \in \mathbb{R}^{N \times N}, B \in \mathbb{R}^{N \times H_{\text {in }}}, C \in \mathbb{R}^{H_{\text {out }} \times N}, D \in \mathbb{R}^{H_{\text {out }} \times H_{\text {in}}}$ are learnable parameters. With special initializations and parameterizations of the parameters~\cite{gu2021efficiently, orvieto2023resurrecting}, LRUs achieve strong performance, as well as enjoy fast inference speed due to the nature of RNNs.
In addition, to achieve efficient parallelizable training, LRUs utilize complex diagonal matrix $A$~\cite{orvieto2023resurrecting}, allowing for a highly parallelizable unrolling to compute the hidden representation $h_k =\sum_{j=0}^{k-1} A^j B x_{k-j}$ with $h_0 = B x_0$. By leveraging eigendecomposition of the complex diagonalizable $A$, we can write $A = P \Lambda P^{-1}$, where $P \in \mathbb{C}^{N \times N}$ is an invertible matrix, and $\Lambda = \operatorname{diag}\left(\lambda_1, \lambda_2, \ldots, \lambda_N\right) \in \mathbb{C}^{N \times N}$. Therefore, the computation of $A^k$ reduces to $P \Lambda^k P^{-1}$ which becomes element-wise diagonal matrix multiplications and significantly improves the training efficiency.
While LRURec~\cite{yue2024linear} pioneers the use of LRUs for sequential recommendation, its complex-valued hidden representations and learnable parameters are not ideal for this task. As demonstrated in~\cite{gu2023mamba}, these complex elements are beneficial for continuous modalities like audio and video, but not for discrete sequence modeling like recommendations. Additionally, complex LRUs introduce a larger number of parameters compared to real-valued counterparts, potentially hindering performance on smaller datasets. Furthermore, the efficiency constraints of LRUs make the parameters in Equation~\ref{eq:lru} static and independent of input user behaviors, unable to capture user behavior within the interaction sequence, limiting further improvement in recommendation accuracy.
Therefore, we aim to utilize real-valued LRUs with dependency on user behaviors, while maintaining fast training speed.

\section{Methodology}

Accurately capturing user behavior nuances is critical for recommendation systems, but traditional linear recurrent units (LRUs) and LRURec sacrifice their behavior-dependency of the learnable parameters to achieve efficient training, limiting their ability to adapt to diverse user interactions. 
To further improve the recommendation performance, we discard the complex-valued parameters and training scheme of traditional LRUs, enabling behavior-dependent modeling. We propose a novel gated behavior-dependent recurrent model, \ours, that tackles the limitations through three key innovations: (1) a recurrent layer with gating mechanisms selectively controlling the relevant input behavior to LRUs, (2) behavior-dependent linear recurrent units (BD-LRUs) that adaptively learn important user behaviors, and (3) efficient hardware-aware parallel scan algorithm enabling parallelizable BD-LRU training on GPUs to compensate for the discarded traditional training scheme. This section introduces the overall architecture and then delves into the key innovations of \ours.

\subsection{Overview}

In this section, we present the overall architecture of our proposed \ours model. As illustrated in Figure~\ref{fig:overview}, it comprises an embedding layer, followed by stacked behavior modeling layers, and a final prediction layer. 
Our behavior modeling layer consists of a gated recurrent layer and a feedforward layer, along with dropout~\cite{srivastava2014dropout}, layer normalization~\cite{ba2016layer}, and residual connections~\cite{he2016deep}.

\noindent \textbf{Gated Recurrent Layer.} 
To further enhance the recommendation performance, our gated recurrent layer incorporates the gating mechanisms used in LSTM~\cite{hochreiter1997long} and GRU~\cite{chung2014empirical}. The gates regulate the flow of user behavior information inside the core Behavior-Dependent Linear Recurrent Unit (BD-LRU). As outlined in Algorithm~\ref{alg:gated_recurrent}, it begins by expanding the hidden dimension $D$ to $DE$ using linear projections implemented by two fully connected layers with an expansion factor $E$, creating two parallel branches.
In the main branch, a unidirectional causal 1D convolutional layer~\cite{fu2022hungry} is applied to preserve the chronological order of the user interaction sequence. A SiLU activation~\cite{hendrycks2016gaussian, ramachandran2017searching} follows this. Next, we use our proposed behavior-dependent linear recurrent unit, which is the core component of the recurrent layer.
In the secondary branch, the input is passed through a SiLU. The outputs of the two branches are then merged via element-wise multiplication.
Finally, a linear projection reduces the dimensionality of the merged output from $DE$ back to the original $D$ dimensions.
This gated recurrent layer design allows the model to capture intricate patterns and interactions within the sequential user behavioral data while effectively maintaining temporal dependencies through the recurrent BD-LRU component and gating mechanisms.

\begin{algorithm}[t]
\vspace{-1pt}
\caption{Gated Recurrent Layer}
\label{alg:gated_recurrent}
\begin{algorithmic}[1]
    \Statex \textbf{Input:} $H_i:(B, T, D)$ \text{(batch size, sequence length, dimension)}
    \Statex \textbf{Output:} $H_o:(B, T, D)$
    \State $H_x:(B, T, ED) \leftarrow \operatorname{Linear}(H_i)$
    \State $H_z:(B, T, ED) \leftarrow \operatorname{Linear}(H_i)$ \\
    \Comment{Linear projections to expand dimensions}
    \State $H'_x:(B, T, ED) \leftarrow \operatorname{SiLU}(\operatorname{Conv1d}(H_x))$
    \State $H_y:(B, T, ED) \leftarrow \text{BD-LRU}(H'_x)$
    \State $H'_y:(B, T, ED) \leftarrow H_y \otimes \operatorname{SiLU}(H_z)$
    \Comment{Merge two branches}
    \State $H_o:(B, T, D) \leftarrow \operatorname{Linear}(H'_y)$ \\
    \Comment{Linear projection to reduce dimensions}
    \State \textbf{return} $H_o$ 
\end{algorithmic}
\end{algorithm}

\noindent \textbf{Feedforward Layer.} 
In the feedforward layer, a position-wise feed-forward network (PFFN) is employed to enhance the modeling of user actions in the latent space. The feedforward network leverages two fully connected layers with a SiLU activation function to capture intricate patterns and interactions within the user interaction sequences. The PFFN is formulated as:
\begin{equation}
\operatorname{PFFN}(H)= \operatorname{SiLU}\left(H\boldsymbol{W}^{(1)}+\boldsymbol{b}^{(1)}\right)\boldsymbol{W}^{(2)}+\boldsymbol{b}^{(2)},
\end{equation}
where $\boldsymbol{W}^{(1)} \in \mathbb{R}^{D \times 4D}$, $\boldsymbol{W}^{(2)} \in \mathbb{R}^{4D \times D}$, $\boldsymbol{b}^{(1)} \in \mathbb{R}^{4D}$, and $\boldsymbol{b}^{(2)} \in \mathbb{R}^{D}$ are the parameters of two fully connected layers. 
Furthermore, to improve model robustness and mitigate overfitting, we incorporate dropout and layer normalization as well.

\subsection{Behavior-Dependent Linear Recurrent Units}

Traditional Linear Recurrent Units have limitations in capturing the nuances of user behavior over time. As shown in Equation~\ref{eq:linrnn}, the parameters of LRUs are independent of input behaviors from different time steps, which makes the LRUs behavior-independent and limits the recommendation performance. Consequently, LRURec based on LRUs struggles to differentiate between diverse user behaviors within interaction sequences, lacking the adaptability to effectively understand how users interact and their preferences evolve. 

To tackle this issue, we design the Behavior-Dependent Linear Recurrent Unit (BD-LRU), whose parameters are dependent on input behaviors across different time steps. The core idea behind BD-LRU is to selectively learn important user behaviors within a sequence. For example, when a user browses various categories before purchasing a product, the model can focus on the browsing history relevant to the eventual purchase, filtering out irrelevant categories the user might have casually explored. This is achieved through the behavior-dependent parameters, which act like a filter, allowing the model to focus on specific elements within the sequence while downplaying others.

Different from traditional LRUs in Equation~\ref{eq:linrnn}, our behavior-dependent Linear Recurrent Unit updates by:
\begin{equation}
	\label{eq:linrnn}
    h_t = \alpha_t \odot h_{t -1} + \beta_t \odot x_t,
\end{equation}
where $\alpha_t$ and $\beta_t$ should not have non-linear dependencies on the prior hidden state $h_{t-1}$, but depend solely on current input $x_t$. We use $\odot$ to denote element-wise multiplication, and $x_t$ represents the input latent representation at time step $t$.

Additionally, to achieve behavior-dependency and control the influence of the past and current input behaviors, we introduce the behavior-dependent recurrence and input gates, given by:
\begin{equation}
    r_t = \sigma (\mathbf W_r x_t + \mathbf b_r),
\end{equation}
\begin{equation}
    i_t = \sigma (\mathbf W_i x_t + \mathbf b_i),
\end{equation}
where $\sigma$ denotes the sigmoid function, normalizing the output of each gate to $[0, 1]$ for reasonable gating semantics.
Both the recurrence gate and the input gate are dependent on the input $x_t$ and independent of the prior hidden state $h_{t-1}$. These gate designs make the parameters dependent on the input of different user behaviors and interactions.
We then follow the design in~\cite{orvieto2023resurrecting, de2024griffin} to calculate the scaling factors:
\begin{equation}
\begin{aligned}
\alpha_t &= \exp(-\operatorname{softplus}(\Lambda) \odot r_t), \\
\beta_t &= \sqrt{1 - \alpha_t^2} \odot i_t, \
\end{aligned}
\end{equation}
where $\Lambda$ is a learnable parameter for the recurrent gate, with the exponential function of a negative value, $\alpha_t$ is guaranteed to be in $[0, 1]$, and the square root term ensures the same for $\beta_t \in [0, 1]$. This formulation ensures the stability of the linear recurrence unit.

For the custom initialization scheme of $\Lambda$, we follow the empirical setting from LRU~\cite{orvieto2023resurrecting} that ensures $\exp(-\operatorname{softplus}(\Lambda))$ is uniformly distributed in $[0.9, 0.999]$. This custom initialization helps yield better performance and stable training as shown in ~\cite{orvieto2023resurrecting}.
The input gate $i_t$ can filter the input latent $x_t$. The recurrence state can mitigate the influence of irrelevant or noisy inputs and selectively focus on relevant information from the inputs and historical behaviors.
Although linear recurrence values can depend only linearly on previous elements, the stacking of linear recurrent layers separated by non-linearities allows for a non-linear dependence on the past. In this sense, the non-linear depth of a linear recurrent network is the number of layers rather than the sequence length. Stacking gated LRU layers allows for rich non-linear dependence on previous events while still taking advantage of fast parallel sequence evaluation.

\subsection{Hardware-Aware Parallel Scan}

Previous linear recurrent units have efficiency constraints that the learnable parameters need to be independent of the inputs of different time steps to perform efficient computation. Moreover, to perform eigendecomposition of the diagonal matrix $A$ in Equation~\ref{eq:linrnn}, the parameters of LRUs must be complex-valued. This complex-value constraint introduces additional parameters without improving performance. Since we have to discard the previous training scheme in LRUs and use real-valued parameters, we employ a hardware-aware parallel scan algorithm to achieve efficient parallelizable training for our BD-LRUs.

\begin{figure}[ht]
    \vspace{-0.1in}
    \centering
    \includegraphics[width=0.33\textwidth]{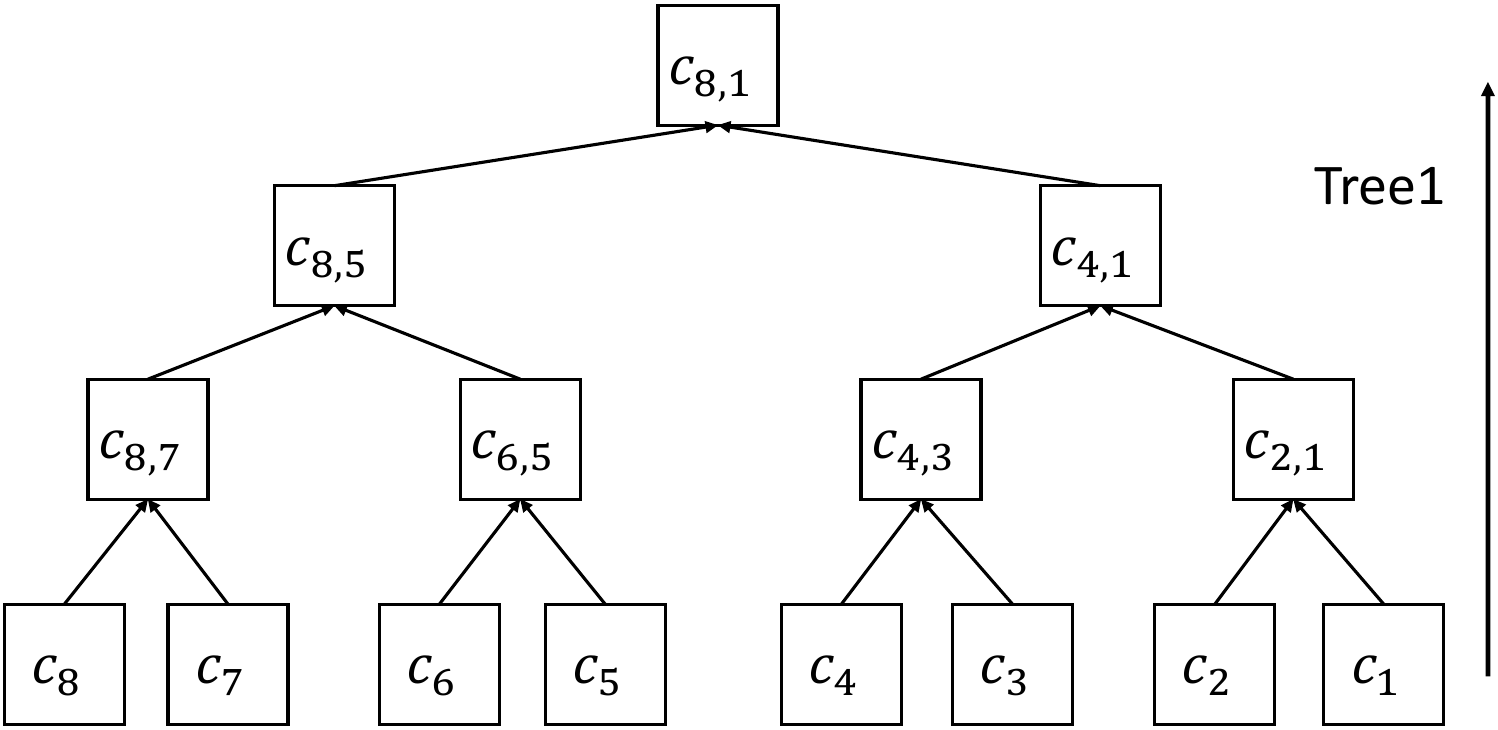}
    \includegraphics[width=0.35\textwidth]{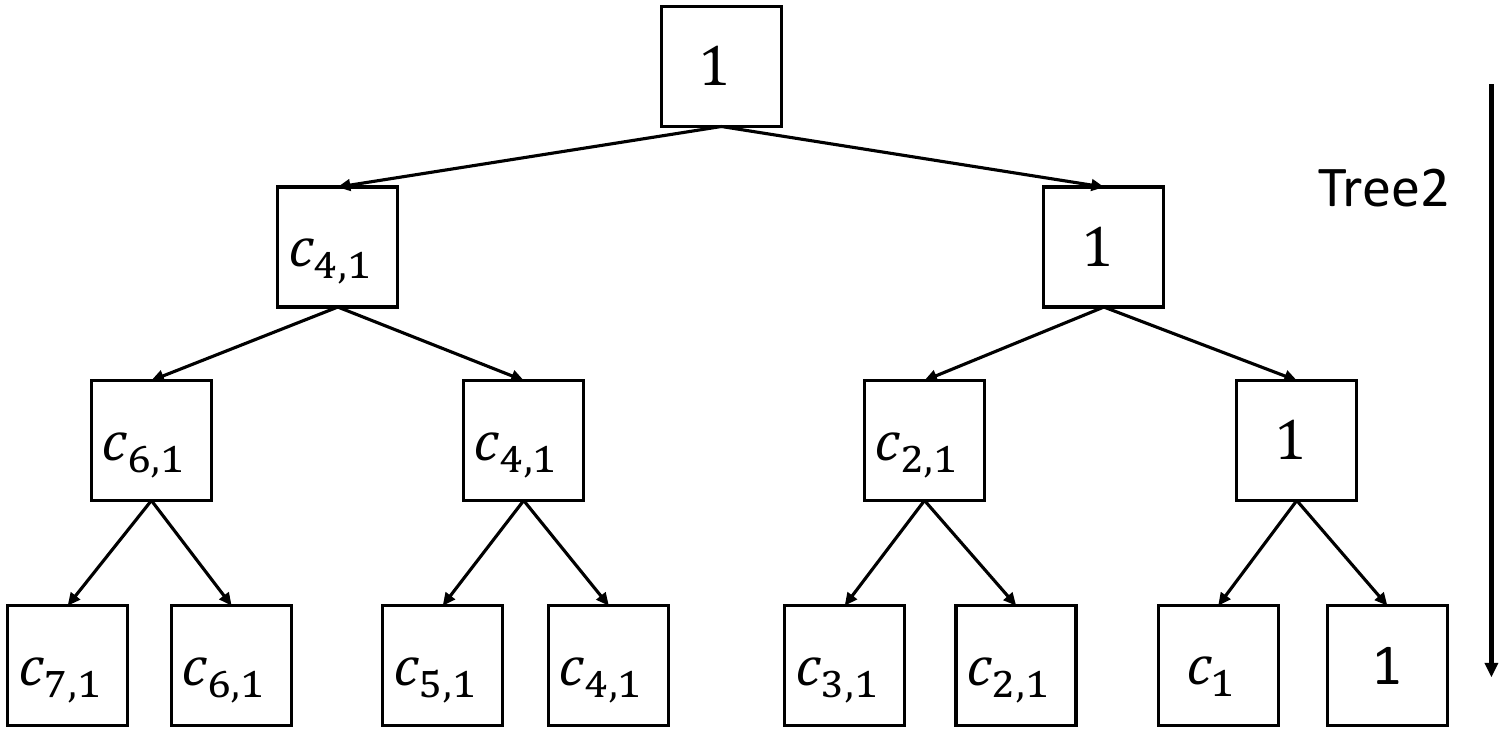}
    \vspace{-0.1in}
    \caption{Parallel scan for computing hidden representations.}
    \vspace{-0.1in}
    \label{fig:scan}
\end{figure}

Since deep learning accelerators like GPUs are only optimized for operators like matrix multiplications and convolutions, we develop an efficient custom CUDA kernel for our BD-LRUs, implemented in OpenAI's Triton~\cite{tillet2019triton}, which is a language and compiler for GPU programming and parallel computing. Our hardware-aware algorithm is based on scans~\cite{blelloch1990prefix, martin2017parallelizing, smith2022simplified}. It has been shown in~\cite{ladner1980parallel} that there exist algorithms for parallel scans when $\oplus$ is associative, \ie, $(a \oplus b) \oplus c = a \oplus (b \oplus c)$. Given a binary associative operator $\oplus$ and a sequence of $n$ elements $a_{1:n}$, the scan operation is defined as:
\begin{equation}
	\operatorname{Scan}\left(\oplus, a_{1:n}\right) = [a_1, (a_2 \oplus a_1), \dots, (a_n \oplus \dots \oplus a_2 \oplus a_1)].
	\label{alg:scan}
\end{equation}
The first-order recurrence is defined as Equation~\ref{eq:linrnn}, where both element-wise multiplication and addition are associative operators (\ie, $(a \odot b) \odot c = a \odot (b \odot c)$ and $(a + b) + c = a + (b + c)$). For simplicity, we express it in:
\begin{equation}
h_t=
\begin{cases}
c_1 & \text{  if } t=1 \\
c_t \bullet h_{t-1} & \text{  if } 1 < t \leq T,
\end{cases}
\end{equation}
where $c_t$ is defined as:
\begin{equation}
c_t=
\begin{cases}
\beta'_1 & \text{  if } t=1 \\
(\alpha_t, \beta'_t) & \text{  if } 1 < t \leq T,
\end{cases}
\end{equation}
when $t > 1$, $c_t$ is an ordered pair. We use $\beta'_t$ to denote $\beta_t \odot x_t$ for simplicity. And the binary operator $\bullet$ is defined as:
\begin{equation}
	\begin{aligned}
		c_2 \bullet c_1 &= (\alpha_2, \beta'_2)\bullet \beta'_1 = \alpha_2 \odot \beta'_1 + \beta'_2 \\
		c_{t+1}\bullet c_t &= (\alpha_{t+1}, \beta'_{t+1}) \bullet (\alpha_t, \beta'_t) \\
		&= (\alpha_{t+1} \odot \alpha_t, \alpha_{t+1} \odot \beta'_t + \beta'_{t+1}) \quad (1 < t < T)
	\end{aligned}
\end{equation}
Therefore, we can derive that:
\begin{equation}
	h_t = c_t \bullet \cdots \bullet c_2 \bullet c_1.
\end{equation}
Since the binary operator $\bullet$ has been shown to be associative in~\cite{blelloch1990prefix}, we do not need to compute the hidden representations $H$ sequentially, but use the scan operation in Equation~\ref{alg:scan}:
\begin{equation}
    \begin{aligned}
        H &= [h_1, h_2, \dots, h_T] \\
        &= \operatorname{Scan}(\bullet, c_{1:T}) \\
        &= [c_1, (c_2 \bullet c_1), \dots, (c_T \bullet \dots \bullet c_2 \bullet c_1)].
    \end{aligned}
    \label{eq:h_scan}
\end{equation}
Our parallel scan algorithm can be demonstrated in Figure~\ref{fig:scan} using two perfect binary trees, which is a case of the sequence length $T=8$. For simplicity, $c_{l, r}$ denotes $c_l \bullet c_{l-1} \bullet \dots \bullet c_{r+1} \bullet c_r$. We initialize all the leaves of Tree1 with $c_8, \dots, c_1$ and the root of Tree2 with $1$. After initialization, we perform an up-sweep on Tree1, then a down-sweep on Tree2, updating the nodes according to the following rules:
\begin{equation}
    \begin{aligned}
        \operatorname{Tree1}[v] &= \operatorname{Tree1}[L[v]] \bullet \operatorname{Tree1}[R[v]] \\
        \operatorname{Tree2}[R[v]] &= \operatorname{Tree2}[v] \\
        \operatorname{Tree2}[L[v]] &= \operatorname{Tree1}[R[v]] \bullet \operatorname{Tree2}[v]
    \end{aligned}
\end{equation}
where $v$ denotes a node in Tree1 or Tree2, $L[v]$ denotes the left child, and $R[v]$ denotes the right child of $v$. After sweeping the binary trees, the hidden representations $H$ in Equation~\ref{eq:h_scan} are computed efficiently. In contrast to serial scan, computing the hidden representations in a sequential manner, with each step depending on the result of the previous step, parallel scan enables parallelizable computation across time steps, significantly enhancing efficiency.

\noindent \textbf{Embedding Padding.}
In real-world scenarios, the lengths of user behavior sequences can vary significantly. We often need to pad these sequences to achieve a uniform length for efficient and consistent processing. Moreover, our BD-LRUs with parallel scans require sequence lengths to be powers of two to build perfect binary trees for parallelizable training. To address this, a dynamic padding strategy should be employed, which involves determining the maximum sequence length, finding the nearest power of two greater than or equal to it, and padding all sequences to that length.
We deviate from the conventional approach of sequence padding and instead leverage embedding padding. Since parallel scans are exclusively performed within the core BD-LRUs, for the remaining convolutional and fully connected layers, padded sequence lengths introduce unnecessary computational costs.
Therefore, strategically apply embedding padding only to the hidden representations $H$ immediately before each BD-LRU, and after the BD-LRU, we truncate the hidden representations $H$ to the original length. This approach proves more efficient compared to sequence padding. Notably, we employ left-padding with zeros to preserve the chronological information within the behavior embeddings.

\vspace{1ex} \noindent \textbf{Algorithm Efficiency.}
We implement a custom CUDA kernel for the hardware-aware parallel scan algorithm to achieve high efficiency for GPU acceleration. The kernel unlocks parallel execution for training BD-LRUs on GPUs. This significantly reduces the time complexity by leveraging multiple GPU processors simultaneously. Furthermore, by employing embedding padding, we further optimize the computation, avoiding unnecessary operations and improving the overall efficiency of the algorithm.

\begin{table*}[]
\setlength{\tabcolsep}{0.5em}
    \centering
    \vspace{-7pt}
    \caption{Performance comparison on five datasets. The best results are highlighted in bold, and the second-best are underlined. The relative improvements in comparison to the best baselines are indicated as Improv. The symbol $*$ indicates a statistically significant improvement of RecBLR over the best baseline with $p$-value < 0.01.} 
    \vspace{-10pt}
    \label{tab:overall}
    \begin{center}
        \begin{tabular}{l l c c c c c c c c c c}
        \toprule
        \textbf{Datasets} & \textbf{Metric} & \textbf{FPMC} & \textbf{Caser} & \textbf{NARM} & \textbf{GRU4Rec} & \textbf{SASRec} & \textbf{BERT4Rec} & \textbf{FMLP-Rec} & \textbf{LRURec} & \textbf{\ours} & \textbf{Improv.}\\
        \midrule
        \multirow{6}{*}{\textbf{ML-1M}}
         & HR@10 & 0.1760 & 0.2399 & 0.2836 & 0.2937 & 0.2993 & 0.2950 & 0.2732 & \underline{0.3057} & \textbf{0.3285}$^*$ & 7.46\%\\
         & NDCG@10 & 0.0933 & 0.1283 & 0.1597 & 0.1638 & 0.1692 & 0.1664 & 0.1416 & \underline{0.1772} & \textbf{0.1901}$^*$ & 7.28\%\\
         & MRR@10 & 0.0682 & 0.0944 & 0.1220 & 0.1245 & 0.1294 & 0.1273 & 0.1056 & \underline{0.1380} & \textbf{0.1478}$^*$ & 7.10\%\\
         
         & HR@20 & 0.2584 & 0.3457 & 0.3841 & 0.4013 & 0.4053 & 0.3978 & 0.3861 & \underline{0.4078} & \textbf{0.4336}$^*$ & 6.33\%\\
         & NDCG@20 & 0.1140 & 0.1550 & 0.1851 & 0.1911 & 0.1958 & 0.1922 & 0.1672 & \underline{0.2030} & \textbf{0.2161}$^*$ & 6.45\%\\
         & MRR@20 & 0.0739 & 0.1017 & 0.1290 & 0.1320 & 0.1367 & 0.1343 & 0.1126 & \underline{0.1451} & \textbf{0.1547}$^*$ & 6.62\%\\
         \midrule
        \multirow{6}{*}{\textbf{Gowalla}}
         & HR@10 & 0.0735 & 0.0747 & 0.0983 & 0.0953 & \underline{0.1202} & 0.1033 & 0.1152 & 0.1189 & \textbf{0.1248}$^*$ & 3.83\%\\
         & NDCG@10 & 0.0415 & 0.0438 & 0.0528 & 0.0513 & 0.0565 & 0.0537 & 0.0569 & \underline{0.0574} & \textbf{0.0601}$^*$ & 4.70\%\\
         & MRR@10 & 0.0318 & 0.0335 & 0.0391 & 0.0380 & 0.0370 & 0.0386 & 0.0379 & \underline{0.0387} & \textbf{0.0405}$^*$ & 4.65\%\\
         
         & HR@20 & 0.1067 & 0.1109 & 0.1472 & 0.1415 & \underline{0.1785} & 0.1541 & 0.1726 & 0.1741 & \textbf{0.1837}$^*$ & 2.91\%\\
         & NDCG@20 & 0.0498 & 0.0519 & 0.0651 & 0.0629 & 0.0712 & 0.0664 & \underline{0.0714} & 0.0713 & \textbf{0.0751}$^*$ & 5.18\%\\
         & MRR@20 & 0.0340 & 0.0367 & 0.0424 & 0.0411 & 0.0410 & 0.0421 & 0.0421 & \underline{0.0425} & \textbf{0.0445}$^*$ & 4.71\%\\
          \midrule
        \multirow{6}{*}{\textbf{Steam}} 
         & HR@10 & 0.1171 & 0.1369 & 0.1315 & 0.1327 & 0.1391 & 0.1366 & 0.1305 & \underline{0.1397} & \textbf{0.1421}$^*$ & 1.72\%\\
         & NDCG@10 & 0.0591 & 0.0675 & 0.0664 & 0.0698 & 0.0705 & 0.0700 & 0.0673 & \underline{0.0710} & \textbf{0.0745}$^*$ & 4.93\%\\
         & MRR@10 & 0.0417 & 0.0468 & 0.0470 & 0.0509 & \underline{0.0511} & 0.0510 & 0.0484 & 0.0504 & \textbf{0.0544}$^*$ & 6.46\%\\
         
         & HR@20 & 0.1670 & 0.2062 & 0.1945 & 0.2004 & 0.2026 & 0.2016 & 0.1919 & \underline{0.2054} & \textbf{0.2116}$^*$ & 3.02\%\\
         & NDCG@20 & 0.0716 & 0.0850 & 0.0823 & 0.0869 & 0.0873 & 0.0871 & 0.0828 & \underline{0.0876} & \textbf{0.0923}$^*$ & 5.37\%\\
         & MRR@20 & 0.0450 & 0.0516 & 0.0513 & 0.0556 & \underline{0.0557} & 0.0554 & 0.0526 & 0.0550 & \textbf{0.0593}$^*$ & 6.46\%\\
          \midrule
        \multirow{6}{*}{\textbf{Beauty}} 
         & HR@10 & 0.0559 & 0.0582 & 0.0627 & 0.0606 & \underline{0.0829} & 0.0780 & 0.0798 & 0.0759 & \textbf{0.0881}$^*$ & 6.27\%\\
         & NDCG@10 & 0.0309 & 0.0318 & 0.0333 & 0.0332 & \underline{0.0409} & 0.0399 & 0.0407 & 0.0393 & \textbf{0.0446}$^*$ & 9.05\%\\
         & MRR@10 & 0.0234 & 0.0239 & 0.0247 & 0.0249 & 0.0281 & 0.0262 & \underline{0.0288} & 0.0280 & \textbf{0.0313}$^*$ & 8.68\%\\
         
         & HR@20 & 0.0825 & 0.0885 & 0.0899 & 0.0858 & \underline{0.1163} & 0.1082 & 0.1159 & 0.1107 & \textbf{0.1243}$^*$ & 6.88\%\\
         & NDCG@20 & 0.0376 & 0.0384 & 0.0405 & 0.0395 & 0.0494 & 0.0459 & \underline{0.0497} & 0.0480 & \textbf{0.0538}$^*$ & 8.25\%\\
         & MRR@20 & 0.0252 & 0.0262 & 0.0267 & 0.0266 & 0.0304 & 0.0286 & \underline{0.0312} & 0.0304 & \textbf{0.0338}$^*$ & 8.33\%\\
          \midrule
        \multirow{6}{*}{\textbf{Sports}} 
         & HR@10 & 0.0301 & 0.0310 & 0.0333 & 0.0318 & \underline{0.0477} & 0.0459 & 0.0474 & 0.0464 & \textbf{0.0500}$^*$ & 4.82\%\\
         & NDCG@10 & 0.0157 & 0.0162 & 0.0170 & 0.0164 & 0.0225 & 0.0208 & 0.0229 & \underline{0.0230} & \textbf{0.0238}$^*$ & 3.48\%\\
         & MRR@10 & 0.0114 & 0.0115 & 0.0121 & 0.0117 & 0.0148 & 0.0129 & 0.0154 & \underline{0.0159} & \textbf{0.0161}$^*$ & 1.26\%\\
         
         & HR@20 & 0.0464 & 0.0475 & 0.0519 & 0.0494 & 0.0691 & 0.0685 & \underline{0.0696} & 0.0671 & \textbf{0.0735}$^*$ & 5.60\%\\
         & NDCG@20 & 0.0198 & 0.0206 & 0.0216 & 0.0208 & 0.0282 & 0.0263 & \underline{0.0285} & 0.0283 & \textbf{0.0297}$^*$ & 4.20\%\\
         & MRR@20 & 0.0125 & 0.0127 & 0.0133 & 0.0129 & 0.0163 & 0.0143 & 0.0169 & \underline{0.0173} & \textbf{0.0175}$^*$ & 1.16\%\\
        \bottomrule
        \bottomrule
        \end{tabular}
    \end{center}
    
\end{table*}

\section{Experiments}

\subsection{Experimental Setup}

\noindent \textbf{Datasets.} We choose the following public datasets for experiments: \textbf{MovieLens-1M}~\cite{harper2015movielens} is a widely-used dataset consisting of 1 million movie ratings.
\textbf{Gowalla}~\cite{cho2011friendship} contains location check-ins from users of the Gowalla location-based social network. \textbf{Steam}~\cite{kang2018self} contains reviews from an online video game distribution platform.
\textbf{Amazon-Beauty} and \textbf{Amazon-Sports} ~\cite{mcauley2015image} contain product ratings and reviews for the Beauty and Sports categories from Amazon.
\textbf{XLong}~\cite{ren2019lifelong} is a dataset from the Alibaba e-commerce platform with long user behavior sequences. We sample 5000 users from the original XLong dataset and use the reduced dataset for assessing efficiency and scalability.
We create a user interaction sequence for each user by chronologically sorting their interaction behaviors based on timestamps. Following the approach adopted in previous studies~\cite{kang2018self}, we exclude users and items with fewer than five recorded interactions. Table~\ref{tab:dataset} summarizes the statistics of the datasets after preprocessing.

\begin{table}[ht]
\normalsize
\tabcolsep=5.5pt
  \centering
  \caption{Statistics of the experimented datasets.}
  \vspace{-10pt}
  \label{tab:dataset}
  \resizebox{0.4\textwidth}{!}{%
  \begin{tabular}{lcccc}
    \toprule
    \multicolumn{1}{l}{\textbf{Dataset}} & \multicolumn{1}{c}{\textbf{\# Users}} & \multicolumn{1}{c}{\textbf{\# Items}} & \multicolumn{1}{c}{\textbf{\# Interactions}} & \multicolumn{1}{c}{\textbf{Avg. Length}} \\ 
    \midrule
    ML-1M & $6,040$ & $3,416$ & $999,611$  & $165.5$\\
    Gowalla & $64,115$ & $164,533$ & $2,018,421$ & $31.5$\\
    Steam & $25,389$ & $4,090$ & $328,378$ & $12.9$ \\
    Beauty & $22,363$ & $12,101$ & $198,502$  & $8.9$\\
    Sports & $35,598$ & $18,357$ & $296,337$ & $8.3$ \\
    XLong & $5,000$ & $329,722$ & $66,822,348$ & $785.9$ \\
  \bottomrule
\end{tabular}}
\vspace{-0mm}
\end{table}

\noindent \textbf{Baselines.} \textbf{FPMC}~\cite{rendle2010factorizing} combines Markov chains and matrix factorization for next-item prediction. 
\textbf{Caser}~\cite{tang2018personalized} employs horizontal and vertical convolutions to learn user preferences.
\textbf{GRU4Rec}~\cite{hidasi2015session} utilizes Gated Recurrent Units~\cite{chung2014empirical} to model user interactions and predict the next-item.
\textbf{NARM}~\cite{li2017neural} is an RNN-based model with the attention mechanism using local and global encoders.
\textbf{SASRec}~\cite{kang2018self} is a unidirectional Transformer-based model leveraging self-attention to adaptively capture user preferences from their behavior sequences.
\textbf{BERT4Rec}~\cite{sun2019bert4rec} uses bidirectional Transformer encoders to model user behavior sequences.
\textbf{FMLP-Rec}~\cite{zhou2022filter} is based on multilayer perceptron with learnable filters using Fast Fourier Transform.
\textbf{LRURec}~\cite{yue2024linear} is an RNN-based recommendation model with complex-valued Linear Recurrent Units~\cite{orvieto2023resurrecting}.

\noindent \textbf{Evaluation Metrics.}
We employ three common metrics: Hit Ratio (HR), Normalized Discounted Cumulative Gain (NDCG), and Mean Reciprocal Rank (MRR). For each metric, we consider the top-$K$ recommendation with a cutoff at $K$, denoted as HR@$K$, NDCG@$K$, and MRR@$K$. We evaluate these metrics at $K = 10$ and $K = 20$.

\noindent \textbf{Implementation Details.} 
Our default model architecture utilizes two behavior modeling layers. We adopt Adam optimizer~\cite{kingma2014adam} with a learning rate of 0.001. A batch size of 2048 is used for training, and 4096 for evaluation. For the XLong dataset specifically, we use a smaller batch size of 512 for training and 1024 for evaluation. Training stops early if validation NDCG@10 fails to improve for 10 consecutive epochs. All models leverage an embedding dimension of 64. The maximum sequence length is set based on the average user interaction length: 1000 for XLong, 200 for MovieLens-1M, 100 for Gowalla, and 50 for other datasets. We refer to a commonly used recommendation library RecBole~\cite{zhao2021recbole} for further implementation details and baseline comparisons. 
Our code is available at \url{https://github.com/chengkai-liu/RecBLR}.

\subsection{Overall Performance}

Table~\ref{tab:overall} presents the overall performance of \ours compared to eight baselines on five datasets. According to the experimental results, Transformer-based models, i.e., SASRec and BERT4Rec, demonstrate superior performance compared to CNN-based Caser and traditional RNN-based NARM and GRU4Rec. FMLP-Rec achieves competitive results on smaller, sparser datasets like Amazon-Beauty, and Amazon-Sports, but remains generally weaker than Transformer-based models and LRURec. Since the linear recurrent unit (LRU) is an improved RNN variant, LRU-based LRURec and \ours have significantly better performance than traditional RNN-based models and achieve competitive results with Transformer models. The experimental results also reveal several key insights for our \ours:
\begin{itemize}
[leftmargin=*,noitemsep,topsep=1.5pt]
	\item \ours outperforms all baselines for all metrics and datasets. This demonstrates that our design significantly improves recommendation accuracy for users, regardless of the data platform.
	\item The improvements are more pronounced on larger datasets with richer user-item interactions and more extensive user behavior histories. For instance, the MovieLens-1M dataset, which has dense interactions and relatively long interaction sequences, exhibits the most significant performance gains. In contrast, performance gains are more modest for sparser datasets with shorter sequences, such as Steam and Amazon-sports datasets.
	\item The improvements in the metrics of NDCG and MRR are typically greater than those in HR. This is because NDCG and MRR are rank-aware metrics, while HR is not. Consequently, \ours excels in ranking, leading to more precise recommendations where users' preferred items appear at higher rankings.
\end{itemize}

\begin{table*}[t]
\centering
\caption{Performance comparison of the variants of \ours.}
    \vspace{-0.1in}
\label{tab:ablation}
\begin{tabular}{lcccccc}
\toprule
\multirow{2}{*}{\textbf{Variants}} & \multirow{2}{*}{\textbf{Metric}} & \textbf{ML-1M} & \textbf{Gowalla} & \textbf{Steam} & \textbf{Amazon-Beauty} & \textbf{Amazon-Sports} \\ 
\cmidrule(l){3-7} 
& & HR / NDCG& HR / NDCG& HR / NDCG& HR / NDCG& HR / NDCG \\ 
\midrule
\multirow{2}{*}{Default} & @10 & \textbf{0.3285} / \textbf{0.1901} & \textbf{0.1248} / \textbf{0.0601} & \underline{0.1421} / \underline{0.0745}  & \textbf{0.0881} / \textbf{0.0446} & \textbf{0.0500} / \textbf{0.0238} \\ 
& @20 & \textbf{0.4336} / \textbf{0.2161} & \textbf{0.1837} / \textbf{0.0751} & \underline{0.2116} / \underline{0.0923} & \textbf{0.1243} / \textbf{0.0538} & \textbf{0.0735} / \textbf{0.0297} \\ 
\midrule
\multirow{2}{*}{Single Recurrent Layer} & @10 & 0.3127 / 0.1802 & 0.1240 / 0.0585 & \textbf{0.1434} / 0.0741 & 0.0808 / 0.0403 & \underline{0.0483} / \underline{0.0229} \\ 
& @20 & 0.4227 / 0.2079 & 0.1830 / 0.0734 & 0.2104 / 0.0911 & 0.1169 / 0.0494 & \underline{0.0714} / \underline{0.0287} \\ 
\midrule
\multirow{2}{*}{BD-LRU Only} & @10 & 0.3010 / 0.1695 & 0.1199 / 0.0579  & 0.1392 / 0.0703 & 0.0840 / 0.0405 & 0.0454 / 0.0223 \\ 
& @20 & 0.4007 / 0.1947 & 0.1805 / 0.0732 & 0.2076 / 0.0876 & 0.1212 / 0.0499 & 0.0698 / 0.0285 \\ 
\midrule
\multirow{2}{*}{w/o Conv1D} & @10 & 0.3245 / 0.1855 & \underline{0.1241} / \underline{0.0599} & 0.1418 / \textbf{0.0754} & 0.0843 / 0.0433 & 0.0480 / 0.0228 \\ 
& @20 & 0.4301 / 0.2121 & \underline{0.1835} / \underline{0.0749} & \textbf{0.2117} / \textbf{0.0931} & \underline{0.1211} / \underline{0.0527} & 0.0697 / 0.0284 \\ 
\midrule
\multirow{2}{*}{w/o PFFN} & @10 & 0.3139 / 0.1780 & 0.1209 / 0.0576 & 0.1404 / 0.0728 & 0.0840 / 0.0425 & 0.0480 / 0.0227 \\ 
& @20 & 0.4172 / 0.2042 & 0.1775 / 0.0718 & 0.2066 / 0.0893 & 0.1209 / 0.0518 & 0.0705 / 0.0282 \\ 
\midrule
\multirow{2}{*}{w/o Custom Initialization} & @10 & \underline{0.3263} / \underline{0.1894} & 0.1188 / 0.0572 & 0.1414 / 0.0741 & 0.0834 / 0.0427 & 0.0479 / 0.0223 \\ 
& @20 & \underline{0.4329} / \underline{0.2151} & 0.1744 / 0.0712 & 0.2100 / 0.0915 & 0.1180 / 0.0514 & 0.0703 / 0.0281 \\ 
\midrule
\end{tabular}
\vspace{-0.1in}
\end{table*}

\subsection{Efficiency and Scalability Study}

For a fair comparison, all experiments in this paper are conducted on a single Nvidia A100 GPU with 40GB of memory. We choose the XLong dataset with a sequence length of approximately 800 for scalability and efficiency study. Due to scalability constraints, we only conduct experiments on selected strong baselines SASRec, LRURec, and our \ours\footnote{Training other RNN-based models, BERT4Rec, and FMLP-Rec is computationally impractical due to their slow speeds or GPU memory requirements exceeding the available 40GB.}. The results are shown in Table~\ref{tab:xlong}, Both LRU-based models \ours and LRURec significantly outperform Transformer-based SASRec because the designed parameterizations of LRUs ensure stability and the ability to avoid vanishing gradients and handle long-range dependencies~\cite{tay2020long, orvieto2023resurrecting}. \ours achieves the best performance among the three models by leveraging its BD-LRUs and gating mechanisms. 

\begin{figure*}[ht]
\vspace{-0.1in}
    \centering
    \includegraphics[width=0.9\textwidth]{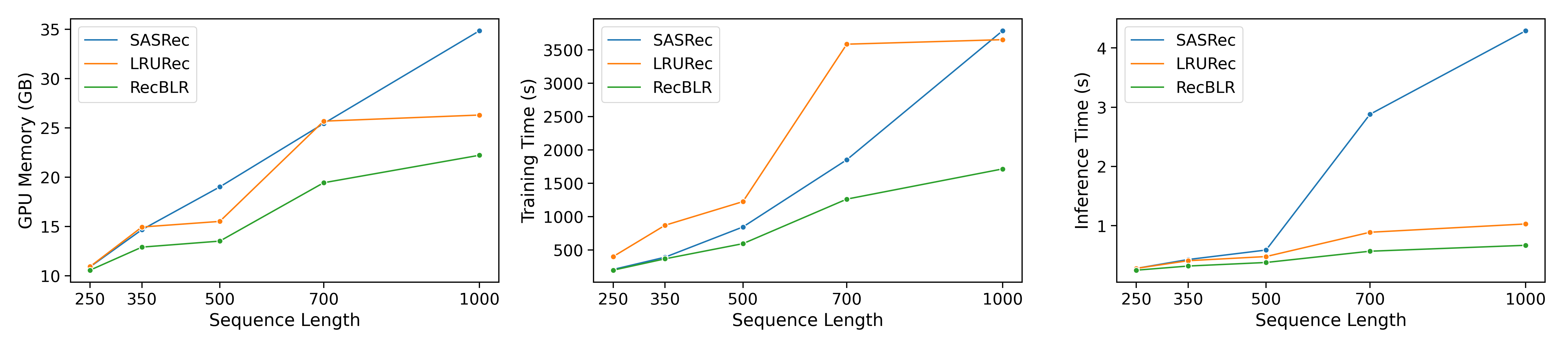}
    \vspace{-0.2in}
    \caption{GPU memory, training time, and inference time per epoch on XLong with different sequence lengths.}
    \vspace{-0.1in}
    \label{fig:sequence_length}
\end{figure*}

\begin{table}[t]
\small
\caption{Performance comparison on XLong with long user behavior sequences.}
    \vspace{-0.1in}
\centering
\begin{tabular}{lcccc}
\toprule
\multirow{2}{*}{\textbf{Models}} & \multicolumn{4}{c}{\textbf{XLong}}                                                     \\ \cmidrule(l){2-5} 
                                  & HR@10 & NDCG@10 & HR@20  & NDCG@20  \\ \midrule
\textbf{SASRec}                   & 0.1946 & 0.1507              & 0.2066 & 0.1538                \\
\textbf{LRURec}                 & 0.2576 &   0.2571         & 0.2586 &      0.2573             \\
\textbf{\ours}                    & \textbf{0.2696} & \textbf{0.2675}       & \textbf{0.2708} & \textbf{0.2678}        \\ \bottomrule
\end{tabular}
\label{tab:xlong}
\end{table}

\vspace{1ex} \noindent \textbf{Efficiency of Different Sequence Lengths.} 
We evaluate the impact on efficiency and scalability across varying maximum sequence lengths for XLong. Figure~\ref{fig:sequence_length} illustrates how the GPU memory cost and training/inference time per epoch escalate as the maximum sequence length increases. \ours excels in both GPU memory usage and training/inference time, particularly for longer sequences. SASRec suffers from the quadratic complexity of its attention mechanisms, leading to rapidly increasing memory usage and training/inference time. While both \ours and LRURec enjoy fast inference speeds due to their recurrent nature, LRURec's training time is considerably longer than \ours, especially when the maximum sequence length is 700. Although LRURec has a parallelizable training scheme, its efficiency suffers due to its incorporation of non-linearity and sequence padding. The parallelizable training of LRURec and \ours requires sequences with lengths that are powers of two. However, the sequence padding (from 700 to 1024) used in LRURec inflates computational cost, whereas \ours employs an embedding padding scheme, incurring fewer additional computations. \ours's efficiency advantage stems from its efficient parallel scan algorithm and embedding padding strategy. This advantage becomes more pronounced with longer sequences, making \ours well-suited for real-world applications involving lifelong user behavior sequences and evolving user interests.

\begin{figure*}[ht]
\vspace{-0.05in}
    \centering
    \includegraphics[width=1\textwidth]{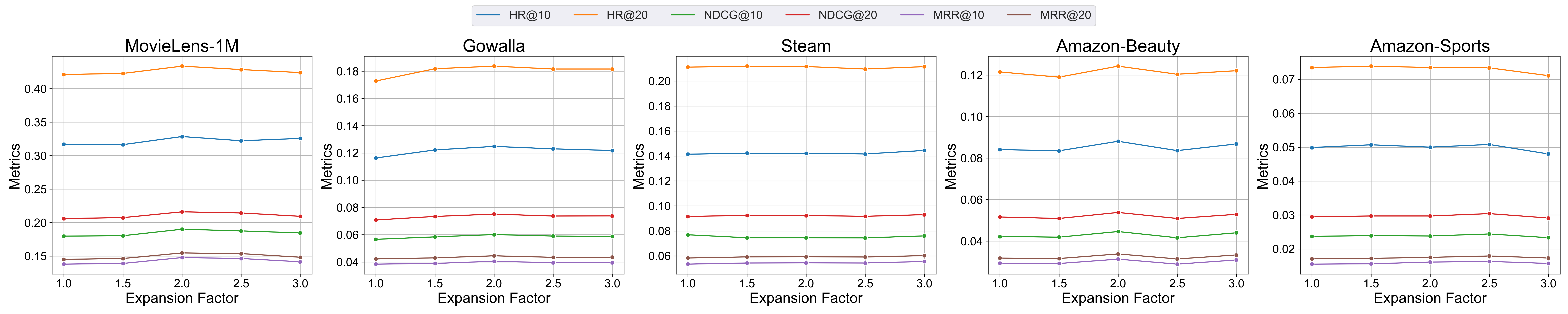}
    \vspace{-0.3in}
    \caption{Influence of the expansion factor on various metrics.}
    \vspace{-0.1in}
    \label{fig:expansion}
\end{figure*}

\begin{figure*}[ht]
    \centering
    \includegraphics[width=1\textwidth]{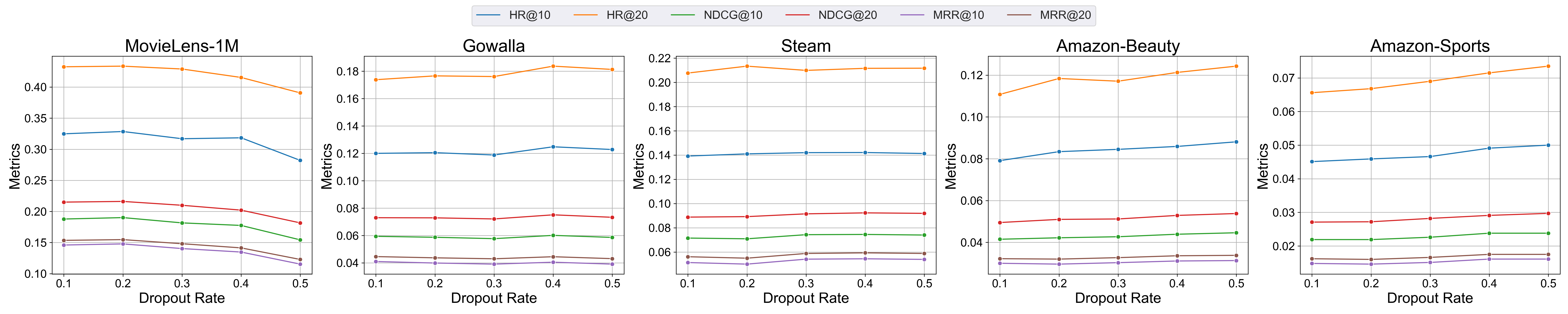}
    \vspace{-0.3in}
    \caption{Influence of the dropout rate on various metrics.}
    \vspace{-0.1in}
    \label{fig:dropout}
\end{figure*}

\vspace{0.2in}
\subsection{Ablation Study}

 To understand how different components contribute to \ours's performance, we conduct an ablation study. We evaluate \ours variants with the following modifications: \textbf{Single Behavior Modeling Layer:} Reduced the behavior modeling layers from two to one. \textbf{BD-LRU Only:} Replaced the gated recurrent layer with a basic Behavior-Dependent Linear Recurrent Unit (BD-LRU), lacking linear projections, gating mechanisms, and causal convolutions. \textbf{Without Conv1D}: Removed the unidirectional causal Conv1D layer from the gated recurrent layer. \textbf{Without PFFN}: Removed the position-wise feedforward network (PFFN). \textbf{Without Custon Initialization}: Replaced the custom LRU parameter initialization with random initialization. Table~\ref{tab:ablation} shows the performance comparison between the default \ours and its variants. We have the following observations and conclusions: \textbf{Generalizability:} The default \ours achieves the best performance across most datasets, demonstrating strong generalization capabilities. \textbf{Number of Behavior Modeling Layers:} While a single behavior modeling layer maintains decent performance, there is a trade-off between the number of layers and computational efficiency. A single layer can significantly improve training and inference speed. \textbf{Importance of Gated Recurrent Layer:} The BD-LRU-only variant, removing linear projections and gating mechanisms in the gated recurrent layer, demonstrates the importance of these components in enhancing the BD-LRU's user behavior modeling capabilities. \textbf{Impact of Conv1D}: Removing the Conv1D layer, which captures temporal dependencies in sequential data, leads to slight performance degradation on most datasets except the Steam dataset, which may be due to the limited size and distribution of Steam. \textbf{Impact of PFFN}: The PFFN contributes to behavior modeling regardless of the underlying operator (LRU or attention mechanisms) in the behavior modeling layer. Removing the PFFN consistently reduces performance. \textbf{LRU Initialization}: The custom initialization scheme commonly used in LRUs ~\cite{orvieto2023resurrecting}, plays a crucial role in enhancing the performance, especially for larger datasets like MovieLens-1M. Using random initialization yields lower performance.

\vspace{1ex} \noindent \textbf{Efficiency Ablation Study.} 
To evaluate the computational efficiency of different scan and padding approaches, we conduct experiments comparing the training time per epoch for serial scan, parallel scan with sequence padding, and our proposed parallel scan with embedding padding. The serial scan computes the hidden representations of linear recurrent units sequentially, as each step depends on the result of the previous step. We employ sequence lengths of 200, 100, and 257 for MovieLens-1M, Gowalla, and XLong datasets, respectively. Due to the constraints of our parallel scan, paddings of 56, 28, and 255 elements are added to the original sequence or embedding lengths of 200, 100, and 257, respectively.
Table~\ref{tab:efficiency} demonstrates that parallel scan significantly reduces training time compared to the traditional serial scan. Additionally, our proposed embedding padding approach yields more time savings compared to sequence padding. This benefit becomes more pronounced with longer extra padded lengths, particularly evident on XLong with an extra padded length of 255, which represents an extreme case for padding.

\begin{table}[t]
\renewcommand\arraystretch{1}
\caption{Training time per epoch on MovieLens-1M, Gowalla, and XLong datasets with different scan and padding schemes.}
    \vspace{-0.1in}
\label{tab:efficiency}
\small
\centering
\resizebox{0.35\textwidth}{!}{
\begin{tabular}{l|c|c|c}
\toprule
\textbf{Scan Algorithm} & \textbf{ML-1M} & \textbf{Gowalla} & \textbf{XLong} \\
\midrule
Serial Scan & 1022s & 908s & 5958s \\
Parallel Scan w/ SP & 75s & 103s & 595s  \\
Parallel Scan w/ EP & 61s & 91s & 263s \\
\bottomrule
\end{tabular}}
\end{table}

\subsection{Hyperparameter Sensitivity}

We conduct the hyperparameter analysis to assess the sensitivity of \ours's to the expansion factor and the dropout rate.

\noindent \textbf{Expansion Factor.}
The expansion factor plays a crucial role in controlling the dimensionality of the hidden representations within the gated recurrent layer. A higher expansion factor can potentially improve the model's ability to capture complex patterns and dependencies in the data, but it may also increase the computational requirements and the risk of overfitting. As shown in Figure~\ref{fig:expansion}, an expansion factor of 2 often yields optimal performance across various datasets. Consequently, for practical purposes, we adopt a default expansion factor of 2 for the experiments.

\vspace{1ex} \noindent \textbf{Dropout Rate.}
We have experimented with various dropout rates across different datasets (Figure~\ref{fig:dropout}). We observe that selecting a higher dropout rate yields better performance for sparser datasets. For instance, the MovieLens-1M dataset, exhibiting relatively low sparsity, achieves optimal performance with a dropout rate of 0.2. Similarly, a dropout rate of 0.4 yields the best results for the Gowalla and Steam datasets. The sparse Amazon-Beauty and Amazon-Sports datasets achieve the highest performance at a dropout rate of 0.5. These observations suggest that considering the dataset's sparsity is crucial when selecting dropout rates.

\section{Related Work}

\noindent \textbf{Sequential Recommendation.}
Sequential recommendation has undergone rapid advancements in recent years. Early approaches Markov Chains (MCs) with matrix factorization~\cite{rendle2010factorizing} to model user preferences based on historical interactions. Recently, deep learning methods have significantly enhanced the expressivity of sequential recommendation systems~\cite{xi2023bird,lin2024can,lin2023map,xi2023towards}. Earlier neural approaches adopting Convolutional Neural Networks~\cite{tang2018personalized} and Recurrent Neural Networks (RNNs)~\cite{hidasi2015session, li2017neural} pioneer the application of neural network in sequence recommendation. Transformer-based architectures have become dominant due to their self-attention mechanism, which captures complex item relationships and improves accuracy. Additionally, the parallelizable nature of attention addresses the efficiency concerns of RNNs. While Transformers offer high accuracy, research continues to optimize inference efficiency. Alternative approaches, such as MLP-based models like FMLP-Rec~\cite{zhou2022filter} and LRURec~\cite{yue2024linear} based on linear recurrent units (LRUs), offer a balance between accuracy and efficiency.

\vspace{1ex} \noindent \textbf{Efficient Sequence Modeling.}
To address the dilemma of performance, training efficiency, and inference efficiency in sequence modeling, various approaches have been proposed such as linear attention, linear RNNs, and state space models (SSMs). Recently, linear RNNs and SSMs have achieved both efficiency and effectiveness. SSMs (\eg,  S4~\cite{gu2021efficiently}, S5~\cite{smith2022simplified}, Mamba~\cite{gu2023mamba}) have been widely adopted as alternatives to traditional RNNs, CNNs, and Transformers in general sequence modeling tasks. Inspired by the success of SSMs, Linear Recurrent Units ~\cite{orvieto2023resurrecting} reinvent RNNs, using linearization, diagonalization, stable exponential parameterization, and custom initialization to create RNN-based models with both strong performance and good efficiency.
Within the recommendation domain, LRURec~\cite{yue2024linear} pioneers the application of linear recurrent units. However, there remains significant potential for exploring and adapting efficient sequence modeling approaches for sequential recommendation tasks.

\section{Conclusion}
In this paper, we propose an effective and efficient sequential recommendation model \ours. By leveraging gating mechanisms and behavior-dependent linear recurrent units, it effectively captures users' evolving interests and behavior patterns, leading to accurate recommendations. The recurrent architecture of \ours enables fast inference, while our designed hardware-aware parallel scan algorithm facilitates efficient training. Extensive experiments on real-world datasets demonstrate \ours' superiority in both recommendation performance and computational efficiency, where \ours outperforms several competitive baselines based on CNNs, RNNs, and Transformers.

\bibliographystyle{ACM-Reference-Format}
\balance
\bibliography{ref}

\end{document}